\documentclass[a4paper,10pt]{edpsci} 
\usepackage{amsmath}
\usepackage{graphicx}
\usepackage{epstopdf}
\usepackage[numbers]{natbib}
\usepackage{hyperref}
\usepackage{url}
\usepackage[switch]{lineno} 

\usepackage{CJKutf8}

\usepackage{arabtex}
\usepackage{utf8}
\setcode{utf8}

\hypersetup{colorlinks=true,citecolor=blue,urlcolor=blue,linkcolor=blue}

\usepackage[ruled]{algorithm2e}
\usepackage{float}
\usepackage{physics}

\newcommand{\e}[1]{{e}^{#1}}
\newcommand{\im}{{i}}

\journalname{}

\articletype{Research article}

\begin{document}

\title{Physical Optics Model to Evaluate Mid-Spatial Frequency Errors on the Point Spread Function}

\titlerunning{This is article title}



\author{Luuk Zonneveld \inst{1}\correspondingauthor{\email{l.zonneveld@tudelft.nl}}
\and
H. Paul Urbach\inst{1}
\and
Aurèle J.L. Adam\inst{1}}




\institute{Department of Imaging Physics, Delft University of Technology, Lorentzweg 1, 2628CJ Delft, The Netherlands}

\abstract{The influence of low-spatial frequency errors of an optical component of an imaging system on the point spread function can be quantified using Zernike polynomials. High-spatial frequency errors cause strong scattering due to which the focused energy is reduced but the shape of the point spread function is mainly preserved. In contrast, the influence of  mid-spatial frequency errors is more difficult to quantify.  Using a scalar diffraction model we evaluate the point-spread function in the presence of such errors and compare the results with experiments.}



\maketitle
\section{Introduction}
    Manufacturing errors  on the surfaces of lenses and mirrors can not be avoided. It is common to distinguish low, mid and high spatial frequency errors.  By definition, low spatial frequency (LSF) errors can, in the exit pupil of the system, accurately be represented by Zernike polynomials up to the 37th order \cite{aikens_specification_2008}. LSF errors can give considerable broadening of the point spread function (PSF), causing deterioration of the image quality. High spatial frequency (HSF) errors of a surface  are of the order of the wavelength. Their influence on the PSF  are studied by statistical models based on the power spectral density of the surface.  High spatial frequency errors often cause considerable light scattering  out of the optical system.  This implies less  intensity in the image, however the sharpness of the image is mainly retained. Surface errors intermediate to the low and high spatial frequencies are the mid-spatial frequency (MSF) errors.  Very high order Zernike polynomials would be needed to capture them, leading often 
    to instabilities \cite{forbes_robust_2010, shakibaei_recursive_2013} and inaccuracies  
    \cite{hosseinimakarem_zernike_2016}. 
    Therefore, the quantification of the influence of MSF errors on the PSF of an imaging system is relatively difficult.

    Several methods to quantify the influence of MSF errors on image quality have been proposed. In \cite{youngworth_simple_2000}  MSF errors are modelled as a perturbation of the surface and statistical methods are applied to evaluate the effect on the PSF.   In \cite{liang_validity_2019}  the first few terms of a WKB expansion of the Helmholtz equation were used to improve the geometric optics model and it was claimed that the model of \cite{youngworth_simple_2000} is accurate only for sufficiently small propagation distances. An alternative method is to incorporate MSF errors in ray tracing software, either retaining the nominal rays or by adapting the ray directions. In \cite{tamkin_theory_2010} MSF errors were propagated to the image plane using the CODE V Beam Synthesis Propagation module and the PSF was calculated in case the MSF were known in the exit pupil.  
    In \cite{stock_description_2017, stock_simulation_2019} special functions are introduced to fit surfaces with MSF errors. 
   For the case of incoherent imaging system several authors have analised the impact of MSF errors on the  OTF \cite{tamkin_theory_mtf_2010, Aryan:19, liang_pupil-difference_2018, liang_effects_2019, demars_pupil-difference_2023, demars_workflow_2024}.

    In the present study the PSF is of interest. To gain insight  we consider here the case that the MSF errors  are limited to a single thin surface where the shape of the surface with errors is incorporated in a transmission function. For simplicity and in agreement with the approach in \cite{liang_validity_2019}, 
    it is assumed that the surface with MSF errors is perfectly imaged by the optical components following it. Perturbations of this image due to  errors in these optical components are assumed to introduce  effects of only second order which are neglected.  
    Assuming that the field in image space is paraxial, the perfect image   of the surface with MSF errors is illuminated by the focussing field without MSF errors and the transmitted field with the MSF errors is finally propagated to the image plane to obtain the PSF. 

    It should be mentioned that several of the assumptions can be relaxed. Instead of assuming a perfect image of the field transmitted by the surface with MSF errors, blurring caused by  LSF surface errors in  the other components following the surface of interest can be taken into account, either by ray tracing or by incorporating them in the transmission functions of these components.  Also several cascaded surfaces with MSF errors can be considered by using Fresnel diffraction integrals to propagate through the optical system. 
    We remark furthermore that the multiplicative transmission function used to describe the surface with the MSF errors can be validated by comparing it to the results obtained with rigorous Maxwell solvers for special, relatively simple  sinusoidal surface shapes. This will be done in a future paper.  

 In the present paper the simulated PSF is validated experimentally. For a couple of relatively simple 1D  MSF topographies, the simulated PSFs are compared with measurements. In the experiment,  the surface shapes are mimicked by programming the corresponding phase profile using a SLM.
    
    The content of the paper is as follows.
 In Section \ref{sec:theory} we derive that the PSF with MSF errors is a convolution of the PSF without these errors (but possibly with LSF errors) and a rescaled 2D Fourier transform of the transmission function of the transmission function that describes the surface with the MSF errors. In Section 3 analytical results are derived for 1D MSF topographies. In Section \ref{sec:ExpSetup} the experimental setup is described and in Section \ref{sec:ResDis} the predictions of the model are compared with the experimental results.

\section{Diffraction model}\label{sec:theory}            
\begin{figure}[!ht]
            \centering
            \includegraphics[width = 8cm, height=5.048cm]{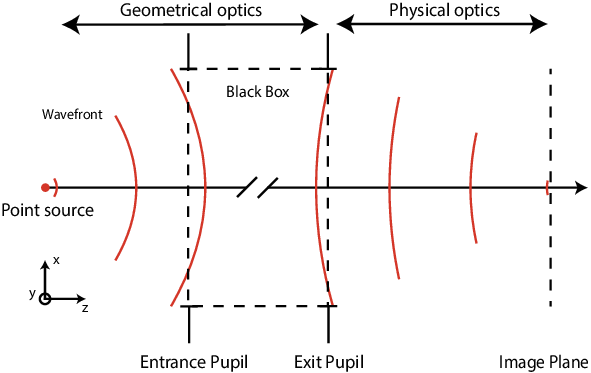}
            \caption{Imaging system  with a point source on the optical axis.}
            \label{fig:blackbox}
        \end{figure}

     An optical imaging system consisting of one or more optical components such as lenses and mirrors, is indicated by "black box" in Fig.~\ref{fig:blackbox}. A point source of which the image is to be determined is on the optical $z$-axis, with $z$ increasing in the direction of the propagation of the light. It is assumed that the numerical aperture of the system is smaller than 0.6 so that scalar paraxial diffraction theory can be used. 

\begin{figure}[!h]
            \centering
            \includegraphics[width=8cm, height=6.223cm]{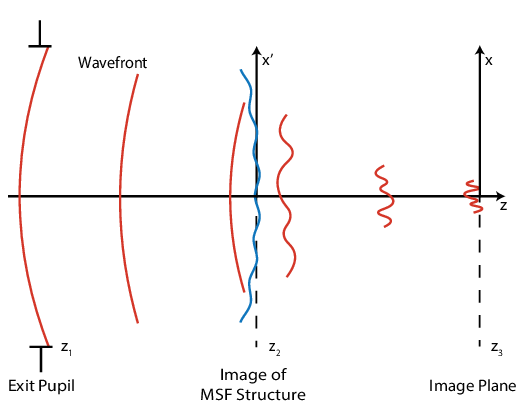}
            \caption{The field in the exit pupil is considered to be unaffected by the MSF structures. It may however contain LSF errors which can be represented by Zernike polynomials. The MSF errors are incorporated in the focused spherical wave from the exit pupil when this wave passes through the image plane of the MSF structure.}
            \label{fig:img_space}
        \end{figure}

    As mentioned in the Introduction, we consider the case that one surface has MSF errors and we assume that the optical components to the right of this surface images the surface perfectly. The perfect image is in the $z=z_2$ plane in Fig. \ref{fig:img_space}. 
   The planes $z=z_1$ and $z= z_3$ correspond to the exit plane and the image plane of the point source, respectively. In the figure $z_1<z_2<z_3$ but the order can be different: the exit plane can for example be to the right of the perfect image plane $z=z_2$ of the surface. The formula below apply also to other orders. 

   We first introduce some notations. $U_z(x,y)$ is a time-harmonic field in the plane with coordinate $z$, which  time dependence is given by  the implicit factor $exp(-i\omega t)$.  For arbitrary $z$ and $z'$ we define the Fresnel propagation integral as
   
\begin{eqnarray}
    U_z(x,y)& = & {\cal P}_{z,z'}(U_{z'})  \label{eq.fresnellabel} \\
    & = & \iint U_{z_1}(x',y')p_{z_1z_2}(x-x', y-y') \text{d}x'\text{d}y',
            \label{eq.fresnelz1z2}
\end{eqnarray}
where

   \begin{equation}\label{eq:propagator}
             p_{z,z'}(\boldsymbol{r}) = \frac{\e{\im k\qty(z-z')}}{\im \lambda \qty(z-z')} \exp( \frac{\im\pi r^2}{\lambda (z-z')}),
        \end{equation}
    with $\boldsymbol{r} = (x, y)$ and $k=\omega/c$,  $c$ the speed of light in the medium between the planes $z$ and $z'$,  and with $\lambda=2\pi/k$ the wavelength.   Note that Eq.~(\ref{eq.fresnelz1z2}) is valid for both  $z_2>z_1$ and $z_2< z_1$, i.e. for both forward and backward propagation.

    For the given point source on the optical axis in object space which image is in the plane $z=z_3$, the field in the exit pupil $z=z_1$  in Fig. \ref{fig:img_space} is given by:
    
        \begin{equation}
            U_{z_1}(\boldsymbol{r}) = P(\boldsymbol{r})\exp(\frac{-ik r^2}{2(z_3-z_1)}),
            \label{eq.Uz1}
        \end{equation} 
    where $P(\boldsymbol{r})$ is the pupil function:
    
        \begin{equation}
            P(\boldsymbol{r}) = \begin{cases}
                \exp(i \Phi(\boldsymbol{r})),  & r\leq a, \\
                0,   & r > a,
            \end{cases}
            \label{eq.pupil}
        \end{equation}
    with $a$  the radius of the exit pupil. The phase function $\Phi$ may contain LSF  aberrations which can be expanded into Zernike polynomials. The perfect (de)-magnified image of the surface with MSF errors is in the plane $z=z_2$. The focused field incident on the $z=z_2$ plane is obtained by applying the Fresnel propagator to the pupil field in Eq.~(\ref{eq.Uz1}): 
    
        \begin{equation}
            U_{z_2} = {\cal P}_{z_1,z_2}(U_{z_1}),
            \label{eq.fresneloperatorz1z2}
        \end{equation}
     This field is perturbed by the the MSF errors:
     
     \begin{equation}
      \tilde{U}_{z_2}(\boldsymbol{r}) = \tau(\boldsymbol{r}) U_{z_2}(\boldsymbol{r}), 
      \label{eq.Utildez2}
     \end{equation}
     where the transmission function $\tau$ is:
     
        \begin{equation}
            \tau(\boldsymbol{r}) = \exp\left[ 
         \frac{2\pi i \Delta n}{\lambda}h(\boldsymbol{r})\right],
         \label{eq.tau}
        \end{equation}
        where $z=h(\boldsymbol{r})$ is the (de)-magnified topography of the MSF errors and  $\Delta n$ is the difference between the refractive index of the optical component with the MSF errors and the surrounding medium. 

The PSF is obtained by propagating the field in Eq.~(\ref{eq.Utildez2}) to the Gaussian image plane $z=z_3$:

        \begin{eqnarray}
           \widetilde{\text{PSF}} =\Tilde{U}_{z_3} & = & {\cal P}_{z_2,z_3}(\Tilde{U}_{z_2}) \nonumber \\
             & = & {\cal P}_{z_2,z_3}( \tau U_{z_2}) \nonumber \\
             & = & {\cal P}_{z_2,z_3}\left[ \tau {\cal P}_{z_1,z_2}(U_{z_1})\right].
\label{eq.Utildez3}
        \end{eqnarray}
    The tilde  indicates that the field contains the effect of the MSF errors. $U_{z_3}$ (without tilde) is the field of the PSF in the image plane without MSF errors, which is the Airy function when there are no aberrations. The expression in Eq.~(\ref{eq.Utildez3}) can be simplified as (see Appendix \ref{appendix:b}):

\begin{eqnarray}
    \widetilde{\text{PSF}}(\boldsymbol{r}) = \frac{\exp( \frac{i \pi  r^2}{\lambda (z_3-z_2)})}{\lambda^2(z_3-z_2)^2} 
            \iint \text{PSF}(\boldsymbol{r}-\boldsymbol{r}') \nonumber \\
            \times\exp( \frac{-i\pi|\boldsymbol{r}-\boldsymbol{r}'|^2}{\lambda (z_3-z_2)})
             \nonumber \\
             \times{\cal F}(\tau)\left(\frac{\boldsymbol{r}'}{\lambda (z_3-z_2)}\right) \, \text{d}x' \text{d}y',
            \label{eq.PSFtilde}
\end{eqnarray}
where ${\cal F}$ is the 2D Fourier transform:

\begin{equation}
    {\cal F}(f)(\boldsymbol{\rho}) = \iint f(\boldsymbol{r}) \exp(-2\pi i \boldsymbol{\rho}\cdot \boldsymbol{r}) \,
     \text{d}x \text{d}y,
      \label{eq.defFFT}
\end{equation}
      where $\boldsymbol{\rho} = (\xi, \eta)$ are spatial frequencies.
      
    Hence, apart form the quadratic phase factors, the PSF in the presence of MSF errors is equal to the PSF without MSF errors convoluted with the Fourier transform of the transmission function $\tau$ representing the MSF errors, evaluated at spatial frequencies $\xi=x/(\lambda (z_3-z_2))$ and $\eta=y/(\lambda(z_3-z_2))$:
    
\begin{equation}
 {\cal F}(\tau)\left(\frac{\boldsymbol{r}}{\lambda (z_3-z_2)} \right).
    \label{eq.convol}
\end{equation}
The function (\ref{eq.convol}) is broader when the distance $z_3-z_2$ between the PSF plane and the plane of the image of the surface with MSF errors is larger and also when
  the MSF errors contain  higher frequencies.
For the special case that there are no other errors except the  MSF errors, the PSF without MSF errors is the Airy disk:

\begin{align}
    \text{PSF}(\boldsymbol{r})= \frac{\exp(ik(z_3-z_1))}{i \lambda (z_3-z_1)} \exp(\frac{ik r^2}{2(z_3-z_1)}) \nonumber \\
    \times a \frac{ J_1\left(2\pi a \frac{ r}{\lambda(z_3-z_1)}\right)}{\frac{ r}{\lambda(z_3-z_1)}}.
    \label{eq.Airy}
\end{align}

The integral in (\ref{eq.PSFtilde}) is a convolution of two functions which is most conveniently computed by  multiplying the Fourier transforms of these  functions and then taking the inverse Fourier transform.
   
    \section{Analytic Formula}
\subsection{1D sinusoidal structure}
 Suppose that the surface error corresponds to a sinusoidal surface perturbation in one direction, say the $x$-direction. Let the (de-magnified) perfect image in the plane $z=z_2$ be given by: 
 
 \begin{equation}
     h(\boldsymbol{r}) = h_1 \cos(2\pi \kappa x + \phi_1).
     \label{eq.surfharm}
 \end{equation}
 where $h_1$ is the amplitude, $\phi_1$ is an off-set and $\kappa$ is the spatial frequency in the plane $z=z_2$. The corresponding transmission function (\ref{eq.tau}) is 
 
 \begin{align}
 \tau(x)&= \exp(i k \Delta n h_1 \cos(2\pi \kappa x + \phi_1))\nonumber \\
 &= \exp(i \bar{h}_1 \cos(2\pi \kappa x) + \phi_1),
 \label{eq.specialtau}
\end{align}
where the dimensionless $\bar{h}_1$ is given by

\begin{equation}
    \bar{h}_1 = k \Delta n h_1.
    \label{eq.barh0}
\end{equation}
With this special  MSF surface, the influence of a particular spatial frequency in the phase error on the PSF can be studied. 
Using the Jacobi-Anger expansion

\begin{equation}
\tau(x) = \sum_{m=-\infty}^\infty i^m J_{m}( \bar{h}_1) \exp(i m (2\pi  \kappa x+\phi_1)),
\end{equation}
we get for spatial frequencies $\boldsymbol{\rho}=(\xi,\eta)$:

\begin{align}
    {\cal F}(\tau)(\boldsymbol{\rho}) = \sum_{m=-\infty}^\infty   J_{m}(\bar{h}_1) \exp(i m (\phi_1+\pi/2)) \nonumber \\
    \times \delta(\xi-m \kappa)\delta(\eta).
    \label{eq.Ftauspecial}
\end{align}
Substitution in Eq.~(\ref{eq.PSFtilde}) gives

\begin{align}
    \widetilde{\text{PSF}}(x,y) =
      \sum_{m=-\infty}^\infty  J_m(\bar{h}_1) \exp(i m(\phi_{jl}+\pi/2)) \,\nonumber \\
      \times \text{PSF}(x-\lambda(z_3-z_2)m\kappa, \, y)\exp(2\pi i m \kappa x)  \nonumber  \\
  \times \exp(-i \pi  \lambda(z_3-z_2) m^2 \kappa^2) 
    \label{eq.PSFharmonic}
\end{align}
It is seen that, in the case of a single frequency MSF phase error, the PSF is a linear superposition of PSFs without MSF errors translated over the distances $\lambda(z_3-z_2)m\kappa$, with $m$ an integer. The relative contribution of the term corresponding to a given value of $m$ depends on the value of $J_m(\bar{h}_1)$.

\subsection{General Case}
Let $h(x,y)$ be a general real function:

\begin{eqnarray}
    h(x,y) =  \int_{-\infty}^\infty \int_{-\infty}^\infty a(\xi,\eta) 
 \cos[2\pi (\xi x+ \eta y) + \phi(\xi,\eta) ] \, \text{d}\xi \text{d}\eta \nonumber \\
    \approx  \sum_{j=-J}^J \sum_{l=-L}^L a_{jl} \Delta \xi \Delta \eta \nonumber \\
    \times \cos[2\pi (j \Delta \xi \,x+ l \Delta \eta \,y) + \phi_{jl}]\nonumber \\
\label{eq.defh}
\end{eqnarray}
where $a_{jl}=a(j\Delta \xi, l \Delta \eta)$ and $\phi_{jl}=\phi(j\Delta\xi, l \Delta\eta)$.
We write

\begin{eqnarray}
  k \Delta n \,h((x,y) = \sum_{j=-J}^J \sum_{l=-L}^L h_{jl} \cos[2\pi (j \kappa_x x+ l \kappa_y y) + \phi_{jl}],\nonumber \\
  \label{eq.rescaledh}
    \end{eqnarray}
where $\kappa_x=\Delta \xi$, $\kappa_y=\Delta \eta$ and $h_{jl}= a_{jl} \Delta \xi \Delta \eta$.  The transmission function is

\begin{eqnarray}
    \tau(x,y) = \exp(i k \Delta n h(x,y)) \nonumber \\
    = \prod_{j=-J}^J\prod_{l=-L}^L \exp(i h_{jl} \cos[2\pi(j \kappa_x x+ l \kappa_y y)+\phi_{jl}])\nonumber \\
    = \prod_{j=-J}^J\prod_{l=-L}^L \sum_{m=-\infty}^\infty J_m(h_{jl}) \nonumber \\\times\exp(i m [(2\pi j \kappa_x x+ 2\pi  l \kappa_y y)+\phi_{lm}+\pi/2]) \nonumber \\
    \label{eq.tau}
\end{eqnarray}
The Fourier transform of the transmission function is:

\begin{eqnarray}
    {\cal F}(\tau)(\xi,\eta) = \prod_{j=-J}^J\prod_{l=-L}^L\sum_{m=-\infty}^\infty  J_m(h_{jl})  \nonumber \\
   \times \exp(i m (\phi_{jl}+\pi/2)) \delta(\xi-m j \kappa_x)\delta(\eta-m l \kappa_y) \nonumber \\
    \label{eq.Ftaugeneral}
\end{eqnarray}
Substitution into Eq.~(\ref{eq.PSFtilde}) implies

\begin{eqnarray}
    \widetilde{\text{PSF}}(x,y) = \prod_{j=-J}^J\prod_{l=-L}^L\sum_{m=-\infty}^\infty  J_m(h_{jl}) \nonumber \\ \times \exp(i m (\phi_{jl}+\pi/2))  \nonumber \\
    \text{PSF}(x-\lambda(z_3-z_2) j m \kappa_x x, y-\lambda(z_3-z_2) l m \kappa_y y) \nonumber \\
    \times \exp(-i \pi \lambda(z_3-z_2)m^2(j^2 \kappa_x^2+l^2\kappa_y^2)) \nonumber \\ \times \exp(-2\pi i m(j \kappa_x x+ l\kappa_y y)). \nonumber \\
    \end{eqnarray}
    
   This formula generalizes Eq.~(\ref{eq.PSFharmonic}) 
   It is only useful when $J$ and $L$ are small. When $J$  or $L$ is not small, the evaluation is too costly and numerical computation should be used instead. 
               
\section{Experimental Method}\label{sec:ExpSetup}
    The model to evaluate the PSF in the presence of MSF errors as described in Section \ref{sec:theory} has been validated with an experimental setup. 
    The MSF errors induce perturbations of the phase of the field. A spatial light modulator (SLM) is used to create these phase perturbations and mimic the MSF errors.
    
    \subsection{Setup}
\begin{figure}[!h]
                \centering
                \includegraphics[width=8cm, height=6.14cm]{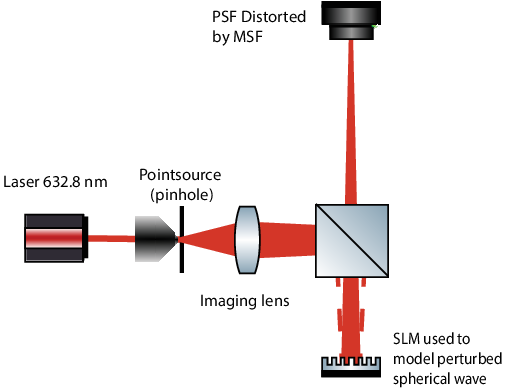}
                \caption{Experimental setup where specific MSF errros are realized in the setup with an SLM. The Resulting PSF is directly imaged onto the camera.}
                \label{fig:setup}
            \end{figure}

        A schematic view of the setup is shown in Fig. \ref{fig:setup}. The light source is a 10~mW polarised HeNe laser emitting at 633~nm. The light intensity is modulated using two polarisers. The second polariser is fixed to ensure that the polarisation direction is  as required by the `\mbox{Holoeye~Pluto~2~VIS-096-D}' SLM.  This SLM is a mirror with a layer of pixelated liquid crystals. Depending on the applied voltage, the orientation of the crystal axis of a pixel changes the orientation of the optical axis and hence the refractive index experienced by the reflected light. This way, in each pixel, the optical path length is adjusted to correspond to the desired phase change. The SLM has a resolution of (1080×1920) square pixels with a side length of 8~$\mu\mbox{m}$. 
        A pinhole of 20~$\mu$m is imaged onto the camera using a Thorlabs best form lens with focal distance $f = 50$~mm. 
        Using a beam splitter after the imaging lens, it is possible to place a reflective SLM in the converging spherical wave originating from the lens. Compared to a transmission SLM, the pixels of reflective SLM devices are often smaller, which is preferred in our setup. The aperture of the lens is limited so that we are as close to diffraction limited performance as possible. This is discussed in Appendix \ref{appendix:ep}. The distance between the lens and the SLM is 18.5~cm, and the distance between the SLM and the camera is 32.7~cm.

    \subsection{Measurements procedure}
        The measurements have first been prepared by eliminating inaccuracies as much as possible. For example, the best form lens was chosen to minimise the chance to add extra MSF structures to the system \cite{messelink_mid-spatial_2018, shultz_effects_2015}. The phase response of the SLM has been calibrated. Also, the surface shape has been measured because SLM displays are typically not optically flat. The surface of the  SLM was characterised by a procedure inspired by the one presented in \cite{liebmann_wavefront_2021}. The method makes use of a Twyman-Green interferometer to determine the height of the SLM surface.

        After characterising the SLM, the next step is to achieve an image of the pinhole on the detector with as few aberrations as possible by aligning the components and introducing an aperture. Components such as the beam splitter, SLM and camera are aligned before adding the lens. Finally, the lens is added and the position is fine-tuned with stages until we cannot observe low spatial aberrations. Important is that we have a large imaging distance to make sure that the Airy disk is large enough and can be measured by the camera. Finally, imaging the PSF is done by turning the first of two polarisers and adjusting the exposure time of the camera so that clear images of the PSF can be taken.

\section{Results}\label{sec:ResDis}
    The model is validated by realizing specific MSF errors with the SLM. In the first subsection, a sinusoidal phase in one direction is realized, for which the PSF can be computed analytically with formula (\ref{eq.PSFharmonic}). In the subsequent subsection, a more complicated phase structure is realized. 

    \subsection{Sinusoidal phase structure}\label{sec:analytical_res}
    For the sinusoidal MSF structure in Eq.~(\ref{eq.specialtau}), and in the absence of low frequency aberrations, the resulting PSF (\ref{eq.PSFharmonic}) is a linear combination of translated Airy disks. These spots can be identified as diffracted orders of the periodic phase structure on the SLM. 
    The $mth$ spot is translated over the distance  $\lambda m\kappa(z_3-z_2)$, where $\kappa$ is the spatial frequency of the sinusoidal structure, and is weighted by the factor $J_m(\bar{h})$ where $\bar{h}$ is the maximum of the phase modulation. Each spot contributes to the final perturbed PSF
    
\begin{figure}[!h]
            \centering
            \includegraphics[width=8cm, height=8cm]{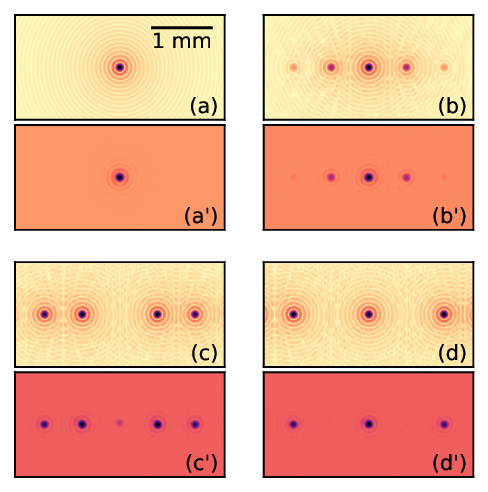} 
            \caption{Four examples of simulation results. In (a) a control image is shown for when the perturbation is set to 0, in (b) a perturbation is shown for $h=\pi/4$, we can see the interference between multiple spots. In (c) and (d) the perturbation is set such that we obtain a zero for $J_0(h)$ and $J_1(h)$ respectively. The simulations are accompanied by the corresponding experiment, indicated with an accent.}
            \label{fig:analytic_res}
        \end{figure}

    Each Bessel function has known zeros depending on the value of the argument, e.g., the function $J_0(h)$ for $h=2.404$ is zero. This means that, according to the model, we can selectively eliminate orders from the image plane. The simulated structure is not related to a specific MSF structure typically found on optical components, but instead is a structure used to verify if the model works as we expect it to. In Fig. \ref{fig:analytic_res} four simulation results are shown using the phase perturbation in Eq.~(\ref{eq.specialtau}). In the simulations, the spatial frequency is kept constant: $\kappa=3.125$~$\text{mm}^{-1}$. The images show the intensity distribution in the normalised image plane, with a gamma correction to improve visibility. In (a) we do not apply a phase perturbation, and in (b) we take a phase perturbation of $h=\pi/4$. In (c) and (d) we make use of the known zeros of the Bessel function to eliminate the 0th and 1st order by setting $h$ to $2.4048$ and $3.8317$ respectfully.
    The simulation results from Fig.~\ref{fig:analytic_res} were generated using parameters measured in the measurement setup, such as imaging distance, aperture size and camera pixel size, in order to get the same PSF in the measurement and simulation. The measurement shown in Fig.~\ref{fig:analytic_res} (a\textquotesingle) shows the PSF that we were able to obtain in the absence of any applied structures while compensating for the surface height variations of the SLM. In Fig.~\ref{fig:analytic_res} (b\textquotesingle), we applied the same phase perturbation on the SLM as we had previously simulated with our model. 
    In Fig.~\ref{fig:analytic_res} (c\textquotesingle) and (d\textquotesingle) we were able to reproduce the removal of the 0th and 1st order. In \ref{fig:analytic_res} (c\textquotesingle) we see that the 0th order does not completely disappear. This is likely because the SLM does not modulate perfectly.

\begin{figure}[!h]
            \centering
            \includegraphics[width=8cm, height=7cm]{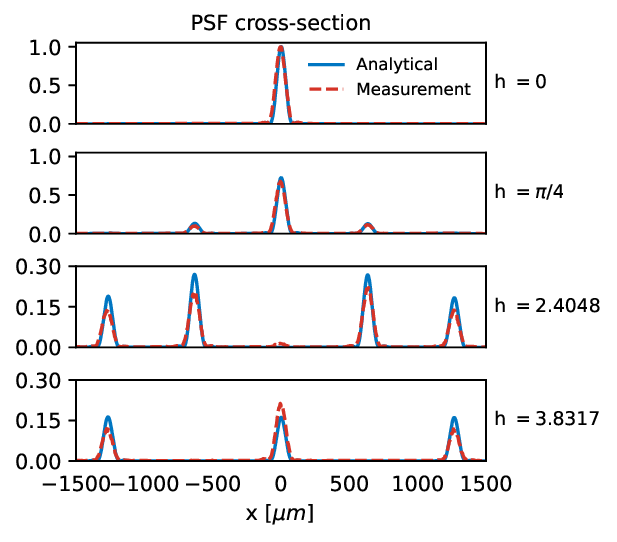} 
            \caption{Overlay of the measurement results through the cross-section of the simulated and measured spots. The pixel size of the detector is $3.45\mu$m}
            \label{fig:analytic_cross}
        \end{figure}

    For a clear comparison, we consider the horizontal cross-sections through the images of both simulated and measured PSF in Fig. \ref{fig:analytic_cross}. The data shown in Fig. \ref{fig:analytic_cross} is normalised to the maximum of the measurement with $h=0$, i.e. for a SLM without phase modulation, but no gamma correction was performed as was done in Fig. \ref{fig:analytic_res}. The cross-sections show that the simulated and measured intensities match well. We can see that for the measurements where we eliminate the orders, we have an increase in the difference between the simulation and measurement. These differences are, however, small. One of the reasons that there is a difference can be the alignment of the structure of the SLM. In the simulations the structure is always perfectly centred but on the SLM this is not possible.

    \subsection{More general MSF structures} \label{sec:numerical_res}
        With the numerical solution, more general MSF structures can be simulated. A one-dimensional phase function consisting of multiple sine functions added is modelled. This way we work towards modelling real MSF structures in a system where we expect a spectrum of frequencies that make up the MSF. 

        We can write the phase function of the perturbation as a series. Each term contains a single frequency and a relative amplitude $h_m$. The structure itself is then finally scaled to again the value $h$
        
            \begin{equation}\label{eq:multiple_sin_perturbation}
                \Phi(x') = h\sum_{m=1}^3 h_m\sin(2 \pi \kappa_m x'), 
            \end{equation}
        We insert the following transfer function into the general expression of Eq.~(\ref{eq.PSFtilde}):
        
            \begin{equation}
                \tau_{msf} = \exp(i\Phi(x')). 
            \end{equation}
            
\begin{figure}[!h]
                \centering
                \includegraphics[width=8cm, height=7cm]{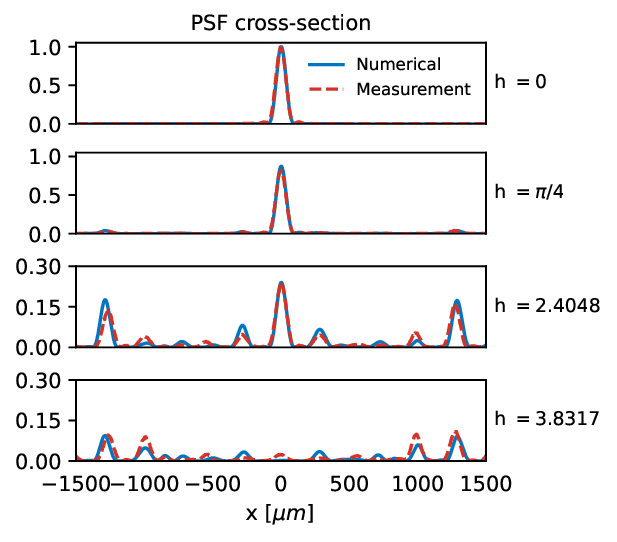}
                \caption{Results for perturbations consisting of multiple frequencies. The frequencies are chosen not to be harmonics. }
                \label{fig:numerical:multiple}
            \end{figure}

        The result obtained using the distortion of the field is shown in Fig. \ref{fig:numerical:multiple}. We used three spatial frequencies, namely: $6.25, 3.57$ and $1.39$~$\text{mm}^{-1}$, and the relative amplitudes of each contribution to the phase perturbation ($h_m$) are $1/2, 1/6$ and $2/6$ respectively, and the values of $h$ are the same as in the results of Fig. \ref{fig:analytic_res}. Just as with the results in Fig. \ref{fig:analytic_res} the graphs are all normalised to their respective maximum amplitude.
        Compared to the results for a single frequency, we observe larger differences between the measurements and the model. By adding multiple sines with different frequencies, the periodicity of the projected structure changes. A shift of the illumination has less impact when a structure of a single frequency is used. When the three sines are added together, we lose periodicity over the SLM display, and any misalignment will alter the results to a higher degree. 
    \subsection{Overlapping orders}
        In the case that the MSF are lower than considered previously, the orders start to overlap and hence interfere. In Figure 7 an example is shown of a single frequency MSF structure where the parameters used are as before except that the frequency of the structure is  $\kappa = 1.786$~$\text{mm}^{-1}$.
            \begin{figure}[!h]
                \centering
                \includegraphics[width=8cm, height=4.8cm]{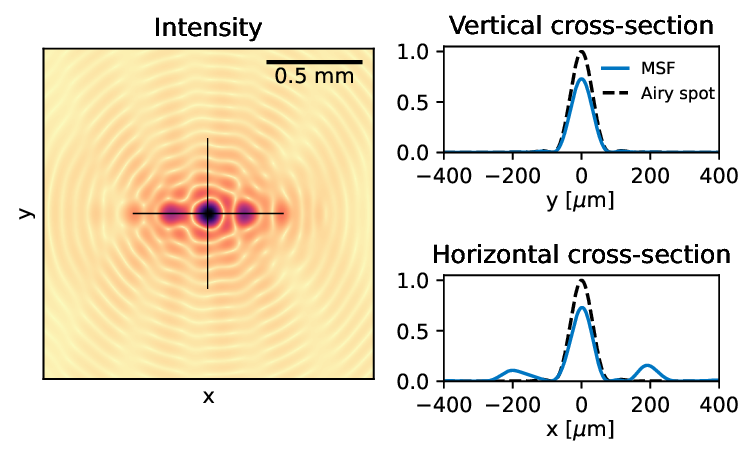}
                \caption{Simulated example where the orders start to overlap. A gamma correction is applied to increase visibility of the PSF. The cross-sections are taken through the centre of the PSF, their extend is indicated on the left with the black lines.}
                \label{fig:overlap}
            \end{figure}
        From the horizontal cross-section it follows that the PSF is deformed due to interference. The shape of the PSF then depends strongly on the value of the amplitude $h_0$.
\section{Discussion and conclusions}
    In this work, we presented a model to simulate the effects of MSF structures on the surfaces of lenses with diffraction integrals. The general mathematical model can calculate the field on the image plane grid. The obtained solution is independent of aberrations. Hence, the method can be used to calculate the PSF under influence of classical aberrations combined with MSF structures.  If the convolution is performed without Fourier transformations, the grid in the plane with the MSF structures can be independently defined from the grid in the image plane. This gives more flexibility in calculating the PSF but as a result the computational complexity is higher. It is however possible to parallelise the computation.
    
    The mathematical model has been verified in an experimental setup, where we observed agreement between the measurements and simulations. However, we should take into consideration that the model has been tested for a low NA. Increasing the NA in the setup is not easy because for higher NA the required resolution of the detector can become too large. In future works, this could be circumvented by working with higher NA setups and phase retrieval. The field can be measured in image space in a more convenient plane and then be propagated to the image plane of the optical system.

    In the setup, we make use of an SLM to quickly test different structures. The advantages are that we can program the SLM with any custom structures, but the limitations are the limited resolution of structures that can be inserted in the model. For further experiments, more optical components can be inserted in the setup and the location of MSF structures can be changed to investigate the effect that it has on the imaging quality of the setup. 

    Some of the errors that we observe when testing the numerical model can most likely be attributed to alignment errors. Small variations in measured intensity compared to the simulations are expected to originate from misalignment of the beam on the SLM. The reason we observe this mostly for the experiments with multiple sinuses added is because the period of the combined frequencies is larger than the SLM. Therefore, the effects are much more pronounced when working with a phase perturbation that is a function of multiple structures compared to a single sinus.

    The results show that the model can be used to estimate the effects of MSF in simple optical systems. However, the model needs to be tested in more complicated optical imaging systems with higher NA, and in the presence of multiple optical surfaces. Also, the extend to which the multiplicative transmission function is a sufficiently accurate representation of the MSF errors should be verified by comparison with a rigorous model. This is the subject of a future paper.

\acknowtext

The authors thank F. Zijp and T. Tukker from ASML, and  G. Vacanti and M.J. Collon from Cosine for the discussions and feedback. 

\funding

This project is co-financed by Holland High Tech with PPP allowance for research and development in the top sector HTSM (TKI HTSM/20.0291).

\conflict

The authors declare that they have no competing interests to report.

\dataavailability

 Data underlying the results presented in this paper are available in Ref. \cite{dataset}.

\authorcontrib

Data curation, software, investigation, visualisation, validation and writing original draft done by L. Zonneveld. Methodology, formal analysis, project administration, conceptualisation, and reviewing and editing done by L. Zonneveld, H.P. Urbach and A.J.L. Adam. Supervision done by A.J.L. Adam and H.P. Urbach. Funding acquisition by H.P. Urbach. 


\appendix

\section{Derivation of Eq. (\ref{eq.PSFtilde}) from  (\ref{eq.Utildez3})}\label{appendix:b}
By taking the Fourier transform of (\ref{eq.fresnelz1z2}) we get

\begin{equation}
    {\cal F}\left[{\cal P}_{z,z'}(U_z)\right]=  {\cal F}(U_z){\cal F}(p_{z,z'}),
    \label{eq.AppendixA1}
\end{equation}
with

\begin{equation}
 {\cal F}(p_{z,z'})(\boldsymbol{\rho}) = \exp(i k (z'-z)) \exp(-i\pi \lambda(z'-z) \rho^2),
 \label{eq.Fpz1z2}  
 \end{equation}
Suppose that for any $U(\boldsymbol{r})$ and $V(\boldsymbol{r})$ we have:

\begin{eqnarray}
{\cal P}_{z,z'}(UV)(\boldsymbol{r}) =  
\frac{\exp\left[\frac{i\pi r^2}{\lambda(z'-z)}\right]}{\lambda^2 (z'-z)^2} 
 \iint  {\cal P}_{z,z'}(U)(\boldsymbol{r}-\boldsymbol{r}') \nonumber \\ \times\exp\left[\frac{-i\pi|\boldsymbol{r}-\boldsymbol{r}'|^2}{\lambda(z'-z)}\right]
 {\cal F}(V)\left[\frac{\boldsymbol{r}'}{\lambda (z'-z}\right] \, \text{d}x' \text{d}y',
\label{eq.propprod}
\end{eqnarray}
  where $\boldsymbol{r}'=(x', y')$.
  Then,
  
   \begin{eqnarray}
       \widetilde{\text{PSF}}(\boldsymbol{r}) = {\cal P}_{z_2,z_3}\left[ \tau {\cal P}_{z_1,z_2}(U_{z_1})\right](\boldsymbol{r}) \nonumber \\
       = \exp\left[ i \pi\frac{r^2}{\lambda(z_3-z_2)}\right]
       \iint 
       {\cal P}_{z_2,z_3}\left[{\cal P}_{z_1,z_2}(U_{z_1})\right](\boldsymbol{r}-\boldsymbol{r}') \nonumber \\
       \times \exp\left[-i\pi \frac{|\boldsymbol{r}-\boldsymbol{r}'|^2}{\lambda(z_3-z_2)}\right]
    {\cal F}(\tau)\left(\frac{\boldsymbol{r'}}{\lambda(z_3-z_2)}\right) \, \text{d}x' \text{d}y' \nonumber \\
       = \exp\left[ i \pi\frac{r^2}{\lambda(z_3-z_2)}\right]\iint 
       \text{PSF}(\boldsymbol{r}-\boldsymbol{r}') \exp\left[-i\pi \frac{|\boldsymbol{r}-\boldsymbol{r}'|^2}{\lambda(z_3-z_2)}\right] \nonumber \\
       \times {\cal F}(\tau)\left(\frac{\boldsymbol{r'}}{\lambda(z_3-z_2)}\right) \, \text{d}x' \text{d}y' \nonumber \\
       \label{eq.tobeproved}
   \end{eqnarray}
where we used that ${\cal P}_{z_2,z_3}\left[{\cal P}_{z_1,z_2}(U_{z_1})\right]=\text{PSF}$. Eq. (\ref{eq.tobeproved}) is identical to (\ref{eq.PSFtilde}). Hence it only remains to derive formula (\ref{eq.propprod}).

Let $\Delta z = z'-z$. We factorize  (\ref{eq.Fpz1z2}) as follows:

\begin{align}
    {\cal F}(p_{z,z'})(\boldsymbol{\rho}) =& e^{ik\Delta z} e^{-i\pi \lambda \Delta z  \rho^2}\nonumber \\
     =&  e^{ik\Delta z} e^{-i\pi \lambda \Delta z |\boldsymbol{\rho}-\boldsymbol{\rho}'|^2}
     e^{i\pi \lambda \Delta z  \rho'^2} 
e^{-2\pi i \lambda\Delta z\boldsymbol{\rho}\cdot\boldsymbol{\rho}'}\nonumber \\
     =&   {\cal F}(p_{z,z'})(\boldsymbol{\rho}-\boldsymbol{\rho}') 
           e^{i \pi \lambda \Delta z \rho'^2} e^{-2\pi i \lambda \Delta z \boldsymbol{\rho}\cdot \boldsymbol{\rho'}}.
    \label{eq.splitting}
\end{align}
Using (\ref{eq.AppendixA1}) with $U_z$ replaced by $UV$, substituting (\ref{eq.splitting}) and using $\boldsymbol{\rho'}=(\xi',\eta')$ gives

\begin{align}
    {\cal F}[{\cal P}_{z,z'}(UV)](\boldsymbol{\rho})  =   
    {\cal F}(U V)(\boldsymbol{\rho})
     {\cal F}(p_{z,z'})(\boldsymbol{\rho})
    \nonumber \\
=    \iint  {\cal F}(U)(\boldsymbol{\rho}-\boldsymbol{\rho}')  
    {\cal F}(p_{z,z'})(\boldsymbol{\rho}-\boldsymbol{\rho}')\nonumber \\
    \times {\cal F}(V)(\boldsymbol{\rho}') e^{i\pi\lambda \Delta z \rho'^2} e^{-2\pi i\lambda \Delta z \boldsymbol{\rho}\cdot\boldsymbol{\rho'}} 
    \text{d}\xi'\text{d}\eta'.  \nonumber \\ 
    = \iint  {\cal F}\left[{\cal P}_{z,z'}(U)\right](\boldsymbol{\rho}-\boldsymbol{\rho}') {\cal F}(V)(\boldsymbol{\rho}') e^{i\pi\lambda \Delta z \rho'^2} \nonumber \\
    \times e^{-2\pi i\lambda \Delta z \boldsymbol{\rho}\cdot\boldsymbol{\rho'}} 
    \text{d}\xi'\text{d}\eta'.  \nonumber \\ 
    \label{eq.FPprodb}
\end{align}
By taking the inverse Fourier transform with respect to  $\boldsymbol{\rho}=(\xi,\eta)$ for fixed $\boldsymbol{\rho'}=(\xi',\eta')$ of the part of the integrand of Eq.~(\ref{eq.FPprodb}) which depends on $\boldsymbol{\rho}$, we obtain the following function of $\boldsymbol{r}=(x,y)$:

\begin{align}
    \iint {\cal F}[{\cal P}_{z,z'}(U)](\boldsymbol{\rho}-\boldsymbol{\rho}') 
    e^{2\pi i(\boldsymbol{r}-\lambda\Delta z \boldsymbol{\rho}')\cdot\boldsymbol{\rho}} \text{d}\xi \text{d}\eta  \nonumber \\
    =  
    \iint {\cal F}[{\cal P}_{z,z'}(U)](\boldsymbol{\tilde{\rho}}) e^{2\pi i(\boldsymbol{r}-\lambda \Delta z \boldsymbol{\rho}')\cdot \boldsymbol{\tilde{\rho}}} \text{d}\tilde{\xi} \text{d}\tilde{\eta} 
    \nonumber \\
    \times   e^{-2\pi i \lambda \Delta z \rho'^2} e^{2\pi i\boldsymbol{r}\cdot \boldsymbol{\rho'}}
    \nonumber \\
  =  {\cal P}_{z,z'}(U)( \boldsymbol{r}-\lambda\Delta z \boldsymbol{\rho}')
    e^{-2\pi i \lambda \Delta z  \rho'^2} e^{2\pi i \boldsymbol{r} \cdot \boldsymbol{\rho}'},
    \label{eq.Finvintegrand}
    \end{align}
where we introduced the change of integration variables $\boldsymbol{\tilde{\rho}} = (\tilde{\xi}, \tilde{\eta}) = \boldsymbol{\rho}-\boldsymbol{\rho}'$. Using (\ref{eq.Finvintegrand}) we find

\begin{eqnarray}
{\cal P}_{z,z'}(UV)(\boldsymbol{r})   = {\cal F}^{-1} \circ {\cal F}[{\cal P}_{z,z'}(UV)](\boldsymbol{r})\nonumber \\ 
 =  \iint {\cal P}_{z,z'}(U)( \boldsymbol{r}-\lambda\Delta z \boldsymbol{\rho}') e^{-\pi i \lambda \Delta z \rho'^2} e^{2\pi i \boldsymbol{r}\cdot \boldsymbol{\rho'}}
      {\cal F}(V)(\boldsymbol{\rho}')  \text{d}\xi' \text{d}\eta'  
     \nonumber \\
=\frac{\exp\left(i\pi \frac{r^2}{\lambda \Delta z}\right)}{(\lambda \Delta z)^2}   \iint {\cal P}_{z,z'}(U)( \boldsymbol{r}-\boldsymbol{r'}) \exp\left(-i \pi\frac{|\boldsymbol{r}-\boldsymbol{r'}|^2}{\lambda \Delta z}\right)\nonumber \\
\times {\cal F}(V)\left(\frac{\boldsymbol{r'}}{\lambda \Delta z}\right) 
   \text{d}\xi' \text{d}\eta'  &\nonumber \\
\end{eqnarray}
This is formula  (\ref{eq.propprod}).

\section{Exit pupil}\label{appendix:ep}
Limiting the aperture of the lens to ensure close to diffraction-limited performance was done by placing an iris almost against the lens surface. The distance was measured to be approximately 3.54~mm. The diameter of the iris is 4~mm, but the effective exit pupil is larger because the Airy disk imaged on the camera was smaller than would be the case for an exit pupil of 4 mm diameter. With ray optics, it is straightforward to determine that the exit pupil is located between the first surface of the lens and the pinhole. The exit pupil is effectively 9~mm in diameter. This is in agreement with the size of the Airy disk imaged on the camera.

\end{document}